\documentclass{article}
\begin{document}
\author{M.P John \\
\\
\baselineskip 5pt\\
International School of Photonics, \\
Cochin University of Science and Technology, \\
Kochi - 682 022, India \\
email:manupj@cusat.ac.in}
\title{Solving a problem in the quantum way }
\maketitle
\baselineskip 9pt
\begin{abstract}
A general quantum algorithm for solving a problem is discussed. The number
of steps required to solve a problem using this method is independent of the
number of cases that has to be considered classically. Hence, it is more
efficient than existing classical algorithms or quantum algorithm, which
requires O($\sqrt{N})$ steps.
\end{abstract}

\baselineskip 12.5pt

\vspace{0.25cm}

\section{ Introduction}

Quantum computation [1,2,3] offers a unique class of algorithms based on
quantum parallelism. It has been shown that certain problems like
factorization [4] and database [5] search can be done more efficiently in a
quantum computer. However designing new quantum algorithms is not an easy
task since our intuitions gained from the world around cannot guide us in
the quantum realm. In this letter, I propose a general quantum algorithm,
which can be applied to all problems where we can classify the numbers in
the initial superposition as solutions non-solutions to the given problem.

\section{The Problem.}

The problem is to find an x such that :

\vspace{0.5cm}

$Cn(x)=0$

\vspace{0.5cm}

$Cn(x)$ can be a mathematical expression or a set of logical statements. It
is known that there are n values of x, which satisfies this condition, and
they may be any number between 0 and $2^k-1$. For example, $Cn(x)$ can be an
n$^{th}$ order polynomial equation, which has n solutions.

\section{ Solving in the quantum way}

The model of the quantum computer I have in mind consists of registers $X$
and $Y$, a quantum processor that is made up of a quantum network whose
action on the registers $X$ and $Y$ is represented by the operator $Uc$. In
addition, a third register Z if needed. A k-bit quantum register can be in
an equally weighted superposition of 0 to $2^k-1$ numbers. We prepare such a
superposition in the X register. We can classify the numbers present in the
supper position into two groups. We also define a new variable y as follows.

\vspace{0.5cm}

{\bf Group 1:} The values of x for which $Cn(x)=0$

\vspace{0.5cm}

{\bf Group2:} The values of x for which $Cn(x)\neq 0$

\vspace{0.5cm}

A new variable y is defined as follows

\vspace{0.5cm}

$y(x)=0$ if $Cn(x)=0$

\vspace{0.2cm}

$y(x)=1$ if $Cn(x)\neq 0$

\vspace{0.5cm}

Our quantum processor can compute y(x) for all values of x. An equally
weighted superposition of the form

\vspace{0.5cm}

$2^{-k/2}\sum\limits_{j=0}^{2^k-1}\mid j\rangle $

\vspace{0.5cm}

is prepared in the $X$ register and the $Y$ register is initially kept in
the state $\mid 0\rangle _y$.Perform the operation

\vspace{0.5cm}

$Uc\mid x\rangle _x\mid 0\rangle _y\longrightarrow \mid x\rangle _x\mid
y\rangle _y$

\vspace{0.5cm}

The state of the system$\mid Q\rangle $after the operation can be
represented as

\vspace{0.5cm}

$\mid Q\rangle =a\sum \mid x_s\rangle _x\mid 1\rangle _y+b\sum \mid
x_{ns}\rangle _x\mid 0\rangle _y$\\
such that $\mid a\mid ^2+\mid b\mid ^2=1$
\vspace{0.5cm}

where $x_s$ are the values which are solutions and $x_{ns}$ are not
solutions to the condition $Cn(x)=0$

\vspace{0.5cm}

Now measure the $Y$ register. If $Y$ register projects to the state $\mid
1\rangle _y$Then due to the entanglement between the $X$ and $Y$ registers,
the $X$ register will contain a superposition of numbers that are solutions.
Measuring the$X$ register will yield a solution. Due to the fact that the
number of solutions in the initial superposition is less compared to the
number of non solutions,  $\mid a\mid^2 <<\mid b\mid ^2.$ Therefore
probability that the $Y$ register projects to the state $\mid 1\rangle _y$
is very less due to the fact that out of the $2^k$entries, only $n$ of them
are the solutions to the condition. Thus the probability to get $\mid
1\rangle _y$ while measuring the $Y$ register is only $n/2^k$ . This means
that most of the times the $Y$ register is projected to the state $\mid
0\rangle _y$. If this is the case, the $X$ register will now contain a
superposition of numbers that are not solutions to the condition $Cn(x)=0$%
.We denote his register by $X^{\sim }$

\vspace{0.75cm}

With this $X^{\sim }$ register, how can we obtain a superposition of
solutions? For this purpose we prepare a superposition of all numbers
between $0$ and $2^k-1$ in a new register $Z$, and by interfering the $%
X^{\sim }$ register and the $Z$ register common modes of the registers get
vanished and now $Z$ register will contain a supperposition of solutions and
measuring $Z$ register, we get one of the solutions.

\vspace{0.5cm}

\subsection{The algorithm can be written as}

\begin{quote}
{\bf \ Step1:}

Prepare an equally weighted superposition of all numbers between $0$ and $%
2^k-1$ in the X register.

\vspace{0.5cm}

{\bf \ Step2: }

Perform the operation $Uc\mid x\rangle _x\mid 0\rangle _y\longrightarrow
\mid x\rangle _x\mid y\rangle _y$

\vspace{0.5cm}

{\bf Step3: }

Measure Y register; If one is obtained measure the X register. If zero is
obtained, perform the interference operation and measure the Z register.
\end{quote}

\section{Conclusion}

\begin{quote}
This General method can be applied to all cases where we can classify the
numbers in the initial superposition as `solution' and `not solution' to the
given condition. For a classical computer, to solve this problem, it has to
check all the numbers between 0 and 2$^k-1$,which requires $2^{k-1}$ steps
on the average. A Quantum computer working on Grover's search algorithm will
take around $2^{k/2}$ steps .The method that we have discussed above,
requires only a finite number of steps, which is independent of the number
of cases that has to be considered classicaly. Hence, for utilizing the advantages of
the quantum parallelism in full, this method or some thing similar to this
has to be developed. The proposed interference step needs further
investigation. I hope that an efficient method for performing this operation
can be found through further research.
\end{quote}

\section{Acknowledgements}

\begin{quote}
I am grateful to Prof M.Sabir, Dept of Physics, Cochin University of Science
and Technology, for giving necessary guidance in the initial stages of the
work. I am also grateful to my colleagues and all those who have given me
encouragement and helpful suggestions.
\end{quote}


\begin{thebibliography}{5}
\bibitem{1}  Andrew, Steane(1998) QuantumComputing Rep.Prog. Phys. 61,117
Preprint: quant-ph/9708022

\bibitem{2}  Spiller T.P{}(1996),Quantum information
processing:cryptography, computation and teleportation, Proc.IEEE 84,1719

\bibitem{3}  M.A.Nielsen and I.L.Chuang (2000) Quantum Computation and
Quantum Information, Cambridge University Press(Cambridge)

\bibitem{4}  Shor,P. W(1994) Polynomial-time algorithms for prime
factorization and discrete logarithms on a quantum computer, in

Proc: 35th Annual Symp. on Foundations of Computer Science,(ed.)
S.Goldwasser,124,IEEE Computer Society Press,

Los Alamitos,CA;preprint quant-ph/9508027

\bibitem{5}  Grover, L. K., (1996) A fast quantum mechanicai algorithm
for database search in Proc.28th Anual ACM Symposium on the Theory of
Computing (STOC),212;quant-ph/9605043
\end{thebibliography}
\end{document}